\documentclass[prl,twocolumn,superscriptaddress,longbibliography]{revtex4-1}
\usepackage{graphicx,subfigure,multirow,makecell}
\usepackage{amsmath,bm}

\usepackage[colorlinks=true, letterpaper=true, pdfstartview=FitV,
  linkcolor=blue, citecolor=blue, urlcolor=blue]{hyperref}

\begin{document}

\title{Berry Phase origin of the planar spin Hall effect}


\author{Haolin Pan}
\affiliation{ICQD, Hefei National Laboratory for Physical Sciences at Microscale, University of Science and Technology of China, Hefei, Anhui 230026, China}
\affiliation{Department of Physics, University of Science and Technology of China,Hefei, Anhui 230026, China}

\author{Dazhi Hou}
\email{dazhi@ustc.edu.cn}
\affiliation{ICQD, Hefei National Laboratory for Physical Sciences at Microscale, University of Science and Technology of China, Hefei, Anhui 230026, China}
\affiliation{Department of Physics, University of Science and Technology of China,Hefei, Anhui 230026, China}

\author{Yang Gao}
\email{ygao87@ustc.edu.cn}
\affiliation{ICQD, Hefei National Laboratory for Physical Sciences at Microscale, University of Science and Technology of China, Hefei, Anhui 230026, China}
\affiliation{Department of Physics, University of Science and Technology of China,Hefei, Anhui 230026, China}

\date{\today}

\begin{abstract}
The Berry phase origin is elaborated for the recent-discovered planar spin Hall effect which features current-induced spin polarization within the plane of the Hall deflection. We unravel a spin-repulsion vector governing the planar spin Hall effect, providing a transparent criteria for identifying intrinsic planar spin Hall materials. Finite spin-repulsion vector is found permitted in 13 crystalline point groups, revealing a big number of unexplored planar spin Hall systems. Our result can be used for the quantitative calculation of the planar spin Hall effect, facilitating further material hunting and experimental investigation.
\end{abstract}

\maketitle

The spin Hall effect (SHE) generates a pure spin current ($j_{\rm s}$) perpendicular to the applied charge current ($j_{\rm c}$) with the spin direction ($s$) orthogonal to both $j_{\rm s}$ and $j_{\rm c}$~\cite{SHEHirsh}. Such a mutually orthogonal orientation between $j_{\rm s}$, $j_{\rm c}$ and $s$ confines the SHE-induced spin polarization within the $x$-$y$ plane for a film sample of $z$ normal direction as illustrated in Fig. 1a. Recent experiment shows that a spin polarization of the $z$ direction~($s_{\rm z}$) can be induced by an in-plane charge current in a material of a mirror plane, e.g. Mo$\rm{Te_2}$, as illustrated in Fig. 2b, which was termed as the planar spin Hall effect~(PSHE)~\cite{PSHENM}. Such a current-induced $s_{\rm z}$, which is out of reach for the spin Hall effect, is highly desirable for device development, because it can switch ferromagnetic layer of perpendicular magnetic anisotropy~(PMA) without involving structural or magnetic field induced asymmetry~\cite{WTe2switching,LiuPRL2012}. Such current-induced $s_{\rm z}$ was also found in some other materials of low-symmetry crystal or magnetic structure, although the terminologies adopted for the phenomenon description are different~\cite{cornelNphys,CornelNC,3mtorque,ChenPRB,Liu2019,Chuang2020,Chen2021}.

However, only a handful of materials have been identified to host the PSHE, and the only clue, the mirror-plane symmetry~\cite{cornelNphys}, still lacks proper microscopic interpretation. To facilitate the material hunting, a microscopic theory of the PSHE is indispensable. The spin conductivity $\sigma_{ijk}$, defined as $J_j^{s_i}=\sigma_{ijk}E_k$, is usually interpreted as a full rank-3 pseudotensor~\cite{Sinova2004,Gradhand2012,Sinova2015}, based on the Kubo formula for the straightforward spin current operator $\hat{J}_j^{s_i}=\frac{1}{2}(\hat{v}_j\hat{s}_i+\hat{s}_i\hat{v}_j)$.  However, it is well known that additional contribution to $\hat{J}_j^{s_i}$ from spin dynamics is needed~\cite{Culcer2004,Sun2005,Shi2006,Nagaosa2008}. Despite several pioneer works towards such direction~\cite{Culcer2004,Sun2005,Shi2006,Sugimoto2006,Chen2006,Nagaosa2008,Zhang2008,Chen2009,Liu2009,Andreas2017,Wesselink2019}, the geometrical origin of the copmlete spin conductivity similar to that in the anomalous Hall effect is still unclear. Such a understanding is tied to the symmetry requirement of the PSHE and can deliver a timing guideline for its further development.

\begin{figure*}
  \centering
  \includegraphics[width=\linewidth]{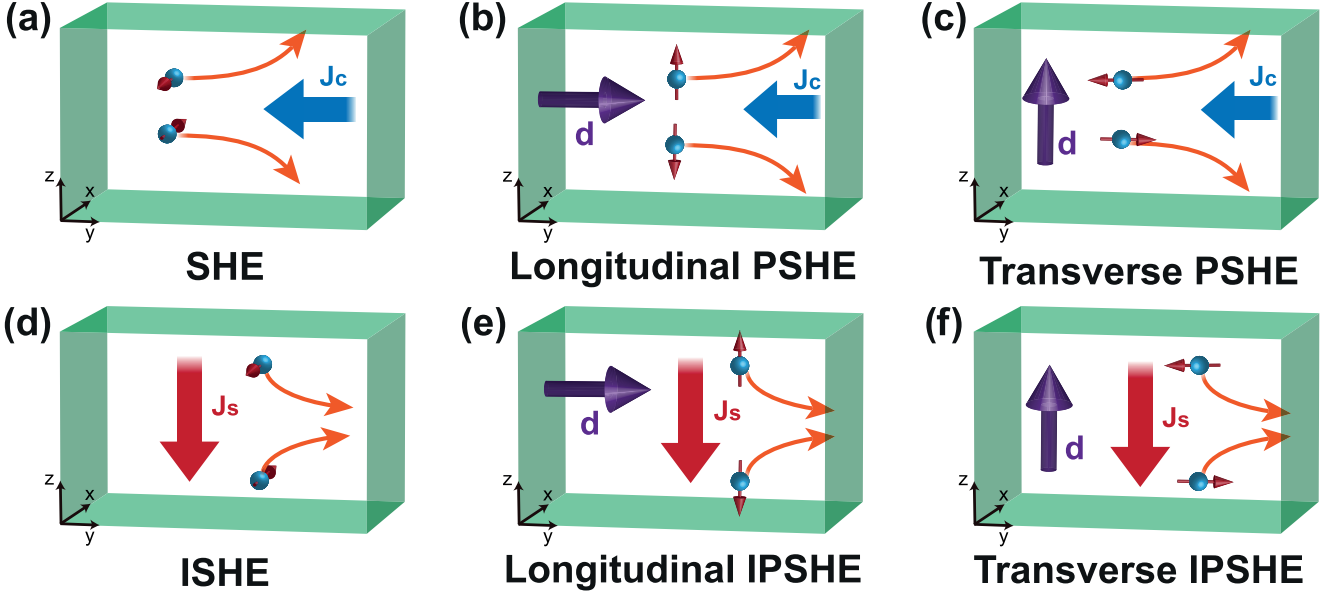}\\
  \caption{Geometrical configuration for the conventional~(a) and planar~(b,c) spin Hall effect; the conventional~(d) and planar~(e,f) inverse spin Hall effect. $\bm d$ is the spin repulsion vector. Depending on $\bm d$ is parallel to the charge or spin flow, we have longitudinal and transverse (inverse) planar spin Hall effect, in which the spin polarization is collinear and perpendicular to the spin current direction, respectively. }\label{fig_fig1}
\end{figure*}



In the letter, we derive the charge current from the spin current injection, which is the inverse effect of the spin current driven by the electric field. Since a proper spin current is subject to the Onsager's relation~\cite{Shi2006}, our theory naturally yields an appropriate spin conductivity $\sigma_{ijk}$ which captures both the  SHE and the  PSHE.

The most striking feature is that the intrinsic part of $\sigma_{ijk}$ is of rank-2, in sharp contrast to the previous knowledge of rank-3. Specifically, $\sigma_{ijk}=-\varepsilon_{jka}D_{ai}$ with $D_{ai}$ the spin-weighted Berry curvature as defined in Eq.~\eqref{eq_conv}. The presence of the Levi-Civita symbol $\epsilon_{jka}$ dictates that the intrinsic contribution to the spin transport is always of the Hall-type, i.e., the spin current always flow perpendicular to the electric field. This is in analogy with the anomalous Hall effect, where the intrinsic contribution to the charge transport is also of the Hall-type.

From the rank-2 feature of the intrinsic spin conductivity, the symmetry requirement of the PSHE can be deduced straightforwardly. Interestingly, the PSHE is determined by a single spin repulsion vector $\bm d$, defined from the antisymmetric part of $D_{ai}$~(see Fig.~\ref{fig_fig1}). We identify 13 crystalline point groups and 30 magnetic point groups permitting finite $\bm d$ and hence allow the PSHE. Several existing experiments for PSHE in materials with mirror symmetries can be well subsumed in our theory and we further predict candidate materials with rotational symmetries as potential platforms of the PSHE. 

{\it Microscopic theory.}---
We now derive the charge current induced by the spin current injection, from which the spin current under the electric field is obtained based on the Onsager's reciprocal relation. By injecting a spin current, an inhomogeneous Zeeman potential $\phi$ is established across a sample, which can be modeled using an inhomogeneous Zeeman field $\bm B^Z(\bm r)$, whose direction and spatial variation represent the spin polarization and flowing direction of the injected spin current, respectively.

To derive the charge current at the order of $\bm \partial_{\bm r}\bm B^Z(\bm r)$, we adopt the semiclassical formalism. We will sketch the main steps and leave the detail in the supplementary materials. Under $\bm B^Z(\bm r)$, the electron motion is governed by the following equations of motion~\cite{Xiao2010}
\begin{align}
\label{eq_rdot}\dot{\bm r}&=\bm \partial_{\bm k} \tilde{\varepsilon}_n-\dot{\bm k}\times \bm \Omega_n-{\bm \Omega}_{\bm k\bm r,n}\cdot \dot{\bm r}\,,\\
\label{eq_kdot} \dot{\bm k}&=-\bm \partial_{\bm r}\phi+{\bm \Omega}_{\bm r\bm k,n}\cdot \dot{\bm k}\,,
\end{align}
where $\bm \Omega_n=-2{\rm Im}\langle \bm \partial_{\bm k}n|\times|\bm \partial_{\bm k}n\rangle$ is the momentum-space Berry curvature in band $n$, $({\bm \Omega}_{\bm k\bm r,n})_{ij}=-2{\rm Im}\langle \partial_{k_{i}}n|\partial_{r_{j}}n\rangle$ is the mixed Berry curvature with $({\bm \Omega}_{\bm k\bm r,n})_{ij}=-({\bm \Omega}_{\bm r\bm k,n})_{ji}$, and $|n\rangle$ is short for $|u_{n\bm k}\rangle$, representing the periodic part of the Bloch wave function. In the velocity equation, $\tilde{\varepsilon}_n$ is the field-corrected $n$-th band energy: $\tilde{\varepsilon}_n=\varepsilon_n+\sum_{ij}\partial_iB_j^Z q_{ij}^S$, where $\varepsilon_n$ is the band energy without the Zeeman field, and the remaining term is the magnetic quadrupolar energy. In the force equation, $\phi=-g_S \sum_i B_i^Z(\bm r) \langle n|s_i|n\rangle$ is the Zeeman potential energy, whose gradient offers the driving force for the elecron motion. As a result, the electron can only reach equilibrium with $\tilde{\varepsilon}_n$ but not $\phi$, leaving the equilibrium Fermi distribution in the form of $f(\tilde{\varepsilon}_n)$.

Another complexity in deriving the transport charge current is that in inhomogeneous systems, the magnetization current generally exists and should be subtracted from the total current. In our case, since the magnetic susceptibility is allowed in any material, an inhomogeneous Zeeman field $\bm B^Z(\bm r)$ always induces a spatially varying orbital magnetization, corresponding to a spatially varying circulation current. After discounting such magnetization current, the transport current reads~\cite{Xiao2010}
\begin{align}\label{eq_cur0}
\bm J=-\sum_n \int \frac{d\bm k}{8\pi^3} \mathcal{D}\dot{r}f_n(\tilde{\varepsilon}_n)+\bm \nabla_{\bm r}\times\sum_n \int \frac{d\bm k}{8\pi^3}\bm \Omega_n g_n\,,
\end{align}
where $\mathcal{D}=1+{\rm Tr}\bm \Omega_{kr,n}$ and $g_n=-k_B T\log[1+\exp[(\mu-\varepsilon_n)/k_BT]$. Since we need the current on the order of $\partial_{\bm r} {\bm B}^Z$, the Berry curvature $\bm \Omega_n$ should be corrected as follows: $\bm \Omega_n\rightarrow \bm \Omega_n+\delta \bm \Omega_n$, where $\delta \bm \Omega_n$ is proportional to $\bm B^Z$. Such correction can be obtained using standard perturbative procedure in the semiclassical formalism.

By plugging all relevant elements into Eq.~\eqref{eq_cur0}, the final expression for the charge current reads~\cite{suppl}
\begin{align}\label{eq_jfinal}
 J^{tr}_i&=- \sum_{jab}\epsilon_{iaj} D_{ab} \partial_j B_b\,,
\end{align}
where $D_{ab}$ is a tensor representing the spin-weighted Berry curvature: $D_{ab}=\int \frac{d\bm k}{8\pi^3} f_n (\Omega_a)_n (s_b)_n$ and $f_n$ is the Fermi distribution function in terms of the origional band spectrum $\varepsilon_n$. This result can be easily implemented in first-principles codes.

Equation~\eqref{eq_jfinal} is one of the main results in this work. Strikingly, the response coefficient involves only a rank-2 tensor $D_{ab}$, in sharp contrast to the rank-3 pseudotensor in previous literatures~\cite{Sinova2004,Gradhand2012,Sinova2015}. To confirm its validity, we also use the linear response theory to derive the charge current and obtain the same result~\cite{suppl}. Moreover, since the relaxation time is not involved, Eq.~\eqref{eq_jfinal} is the intrinsic contribution to the charge current.

The Onsager's relation has been shown to hold for a properly defined spin current by accounting the spin torque~\cite{Shi2006}.  It imposes the following spin conductivity from Eq.~\eqref{eq_jfinal}: $\sigma_{ijk}=-\varepsilon_{jka}D_{ai}$, whose essential element $D_{ai}$ is of rank-2. This is clearly different from the rank-3 spin Hall conductivity used in first-principles calculations in previous literatures. Such a difference is most clearly reflected in the symmetry requirement as shown later.

Due to the Levi-Civita symbol $\epsilon_{iaj}$, the charge current is always normal to the spin current, i.e., only the Hall-type current exists for the intrinsic contribution. This is surprisingly similar to the intrinsic contribution to the anomalous Hall current:  $(s_b)_n$ serves as the `charge' of the driving force and by replacing it with the electric charge $-e$, Eq.~\eqref{eq_jfinal} reduces to the familiar Berry phase contribution to the anomalous Hall conductivity. The geometrical origin of the inverse spin Hall effect is then the interplay of the `charge' and Berry curvature.

Since for the `charge' $(s_b)_n$ is a vector, it is natural to find that there are two types of Hall currents, depending on the relative orientation between the `charge' and the Hall-deflection plane. When $(\bm s)_n$ is normal to the Hall deflection plane~(diagonal part of $D_{ab}$), we get the conventional inverse spin Hall effect; when $(\bm s)_n$ is within the Hall deflection plane~(off-diagonal part of $D_{ab}$), we obtain the planar inverse spin Hall effect. The latter one is our focus in this work.

To prove such classification, we note that as a rank-two tensor, $D_{ab}$ can always be splitted into a symmetric and antisymmetric part: $D^{s,as}=(D\pm D^T)/2$. Since $D^s$ is orthorgonal, it can always be diagonalized. Therefore, we can choose the real space axis to be along the eigen vectors of $D^s$, imposing such diagonalization of $D^s$. The resulting charge and spin current reads
\begin{align}\label{eq_conv}
J_i^{tr}=-\epsilon_{iaj}D_{aa}^s \partial_j B_a\,,
\end{align}
 for a Zeeman field pointing along $a$-th direction and changing along $j$-th direction. We find that the spin polarization and flow direction of the spin current and the charge current flow direction are perpendicular to each other. This is thus the conventional spin Hall effect and its inverse, as shown in Fig.~\ref{fig_fig1}(a).

On the other hand,  the antisymmetric part $D^{as}$ always induces a charge current with the planar configuration, i.e., the PSHE and its inverse. To see this, we define a vector $\bm d$ from $D^{as}$:
\begin{align}\label{eq_gspd}
\bm d=\frac{1}{2}\epsilon_{abc}\hat{e}_aD^{as}_{bc}=\int \frac{d\bm k}{16\pi^3} f_n \bm \Omega_n\times \bm s_n\,.
\end{align}
 Due to the cross product, $\bm d$ is always perpendicular to spin and we thus refer to it as the spin repulsion vector, which plays an essential role in the planar configuration of the current.

In terms of the spin repulsion vector $\bm d$, the charge current reads $J_i^{tr}=J_i^\text{lon}+J_i^\text{tran}$ with
\begin{align}\label{eq_pshe}
J_i^\text{lon}=-d_i \partial_j B_j,\; J_i^\text{trans}=d_j \partial_j B_i\,,
\end{align}
where $i\neq j$. Comparing with the conventional inverse spin Hall effect in Eq.~\eqref{eq_conv}, only two different spatial indices appear, indicating a planar configuration. Moreover, the spin repulsion vector can be either parallel with or perpendicular to the charge current flow, corresponding to the longitudinal and transverse  configuration, respectively, as shown in Fig.~\ref{fig_fig1}(e) and (f).

Using the Onsager relation, we can also write down the electric-field-induced spin current: $
J_i^{s_j,trans}=d_i E_j,\; J_i^{s_i,lon}=-d_j E_j$, with $i\neq j$, as shown in Fig.~\ref{fig_fig1}(b) and (c). One immediately finds that the longitudinal PSHE supports a spin current with parallel spin polarization and flow direction. It can then be used for magnetization reversal of magnetic heterostructures with perpendicular magnetic anisotropy.

{\it Symmetry requirement.}---Although the conventional and planar spin Hall effect shares similar physical pictures, the study of the latter one only starts recently, due to its stringent symmetry requirements compared with the former one.
  Since both $\bm s$ and $\bm \Omega$ transform as an axial vector, the diagonal part $D_{aa}$ is not subject to any symmetry constraint, consistent with the wide existence of the conventional spin Hall effect.

In contrast, to allow the PSHE, the spin repulsion vector $\bm d$ must exist. Since $\bm d$ transforms as an axial vector under point-group operations, it flips sign under mirror operations that contain $\bm d$, but stays the same if the mirror plane is perpendicular to it; it also changes its direction under rotation about an axis perpendicular to it.

\begin{figure}[t]
  \centering
  \includegraphics[width=\columnwidth]{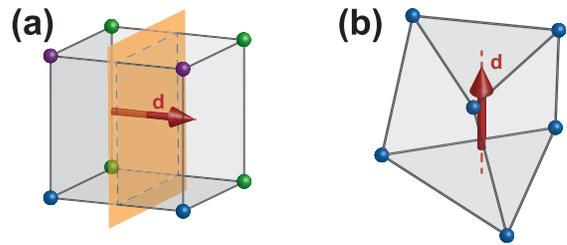}\\
  \caption{The spin repulsion vector in lattices with mirror symmetry~(a) and rotational symmetry~(b). }\label{fig_fig2}
\end{figure}

Based on this analysis, we find that among the 32 crystalline point groups, 13 of them allow an axial vector, including $C_n$, $C_{nh}$ with $n=1,2,3,4,6$, and $S_{2n}$ with $n=1,2,3$. In particular, $\bm d$ is perpendicular to the plane of the mirror symmetry and parallel with the rotational axis of $C_n$, as shown in Fig.~\ref{fig_fig2}.

The only experimental clue for the PSHE up to now is the mirror symmetry, which can be well explained by our theory. As a first example, we consider ${\rm MoTe_2}$~\cite{PSHENM}. The structure of the ${\rm MoTe_2}$ film preserves only a single mirror symmetry with the mirror plane perpendicular to the interface. As a result, a spin repulsion vector $\bm d$ parallel to the interface exists. Based on Fig.~\ref{fig_fig1}(b), when the electric field is applied parallel to $\bm d$, a spin current can flow out of the interface, with a spin magnetization also perpendicular to the interface, as observed in the experiment.

There are also magnetic materials allowing the PSHE. In our theory, $\bm d$ and magnetism can indeed coexist. However, as an axial vector, $\bm d$ is distinct from its fellow quantities, such as spin or magnetic field, in that $\bm d$ stays the same under time reversal operation while spin and magnetic field flip sign. This has profound implication in magnetic materials. For example, in a ferromagnet with the magnetic point group $m^\prime m^\prime m$, the magnetization $\bm M$ is a natural axial vector, which lowers the symmetry of the crystal by breaking any mirror symmetry that contains $\bm M$. Nevertheless, $\bm M$ alone cannot support a nonzero $\bm d$ parallel to it: under the combined operation $\mathcal{M}T$ with any mirror plane containing $\bm M$, $\bm M$ stays the same but $\bm d$ flips sign. This difference excludes a simple ferromagnet as a candidate for the PSHE, unless the additional $\mathcal{M}T$ symmetry is further broken. It means that the intrinsic PSHE cannot be used to explain the current-induced $s_{\rm z}$ which changes its sign when the magnetization flips the direction, e.g. in Co and  $\rm MnAu_2$\cite{Chuang2020,Chen2021}.

We then identify all magnetic point groups that allow $\bm d$ as shown in Table~\ref{tab_mpg}. Interestinly, $\bm d$ can exist in both ferromagnets and antiferromagnets. However, as discussed previously, since $\mathcal{M}T$ is equivalent to $\mathcal{M}$ for $\bm d$, only a single mirror operation may exist in all those groups, which are all perpendicular to the possible rotational axis.

Armed with this knowledge, we can offer new perspectives to existing magnetic systems showing PSHE. A good example is the ${\rm Mn_3 GaN}$ film with the (001) plane as the interface\cite{CornelNC}. Although the crystal symmetry is high, with both a four-fold rotational symmetry and four mirror planes that contain the rotational axis, the antiferromagnetic order efficiently reduces the point-group symmetry without introducing any additional combined symmetry as discussed previously. The remaining mirror symmetry then allows $\bm d$ parallel to the interface, similar to ${\rm MoTe_2}$, and the PSHE then emerges.

\begin{table}
\begin{tabular}{|c|c|c|}
\hline
crystal systems &
 magnetic order &
group notation\\
\hline
\multirow{2}{*}{Triclinic} & F & $1$,$-1$ \\ \cline{2 -3} & AF & ${\rm -1^\prime}$ \\ \hline
\multirow{2}{*}{Monoclinic} & F & $m$, ${ m^\prime}$, 2, ${ 2^\prime}$, $2/m$, ${ 2^\prime/m^\prime}$ \\ \cline{2 -3} & AF & ${ 2/m^\prime}$, ${ 2^\prime /m}$ \\ \hline
\multirow{2}{*}{Tetragonal} & F & 4, 4/m, -4 \\ \cline{2 -3} & AF & ${\rm 4^\prime}$, ${\rm 4^\prime /m}$ ,${\rm 4/m^\prime }$,${\rm 4^\prime/m^\prime }$, ${\rm -4^\prime }$\\ \hline
\multirow{2}{*}{Rhombohedral} & F & 3, -3 \\ \cline{2 -3} & AF & ${\rm -3^\prime }$\\ \hline
\multirow{2}{*}{Hexagonal} & F & 6, 6/m, -6 \\ \cline{2 -3} & AF & ${\rm 6^\prime}$, ${\rm 6^\prime /m}$ ,${\rm 6/m^\prime }$,${\rm 6^\prime/m^\prime }$, ${\rm -6^\prime }$\\
\hline
\end{tabular}
\caption{Magnetic point groups that allow the existence of $\bm d$, with the corresponding magnetic orders~(F or AF). Magnetic point groups that explicitly contain the time reversal symmetry are excluded as they have been discussed in the main text. Rhombic and Cubic crystal systems forbid $\bm d$ and are hence not listed.}\label{tab_mpg}
\end{table}

Both ${\rm MoTe_2}$ and ${\rm Mn_3 GaN}$ rely on the interface to lower the point-group symmetry of the bulk. In comparison, the collinear antiferromagnetic IrMn allows $\bm d$ in the single crystal form. IrMn is a cubic crystal, and the antiferromagnetic ordering lowers the point-group symmetry to $2/m1^\prime$. $\bm d$ is therefore allowed along the $[\bar{1}10]$ direction.

Special attention should be paid to the non-collinear antiferromagnet ${\rm Mn_3Ir}$. The magnetic order lowers the crystal point-group symmetry to $-3 m^\prime$. Since the rotational axis is contained in the mirror plane, $\bm d$ is not allowed in such a system. Therefore, no intrinsic PSHE can exist in the bulk, contrary to the first-principle calculations~\cite{Zhang2017,Liu2019}. This discrepancy may arise from the fact that the incomplete spin current operator is used in the DFT calculations and it points the currently observed spin accumulation with spin direction perpendicular to the interface to other origins~\cite{Liu2019}, such as the magnetoelectric effect or heterostructures with special interfaces.

As discussed above, material systems based on the mirror symmetry can generally be explained by our theory. However, based on our symmetry analysis, we find that the spin reuplusion vector $\bm d$ and the PSHE consistent with the rotational symmetry are largely unexplored. This points alternative direction for future study of the PSHE. Candidates include ${\rm Al_2CdS_4}$~(point group $-4$), ${\rm VAu_4}$~(point group $4/m$), and ${\rm BiI_3}$~(point group $-3$).

The microscopic origin of the PSHE consistent with the rotational symmetry can be further illuminated. It is well established that the key element for spin Hall effect is the spin-orbital coupling, which is highly sensitive to the pont-gorup symmetry. In crystals with the rotaional symmetry, a minimal model for the conduction electron with a symmetry-allowed spin-orbital couplling can be put in the following form
\begin{align}\label{eq_soc}
\hat{H}=\frac{k^2}{2m}+\lambda (k_x \sigma_y-k_y\sigma_x+k_z\sigma_z)\,,
\end{align}
where the first term is the kinetic energy and the remaining terms are the spin-orbital coupling in three dimension with $\bm \sigma$ being the Pauli matrix for spin. Such a model is chiral and respects the rotaional symmetry about the $z$-axis~\cite{Sun2015}. 

For a general chemical potential $\mu$, the spin repulsion vector at zero temperature can be evaluated. The result reads $\bm d=(0,0,d_0)$ with~(we take $e$, $\hbar$, and lattice constant $a$ to be unity)
\begin{align}
d_0=\frac{\sqrt{2m\mu +m^2\lambda^2}}{6\pi^2}\,.
\end{align}
This is consistent with the symmetry analysis and the resulting PSHE increases with the chemical potential, as it is a Fermi-sea effect.The model in Eq.~\eqref{eq_soc} then illustrates the type of spin-orbital coupling needed for the PSHE with rotaional symmetry.

{\it Discussions.}---We have shown that the intrinsic part of the spin conductivity is of rank-two and the resulting PSHE is determined by a single polar vector $\bm d$. We shall note that the extrinsic contribution to the spin conductivity has been ignored, which may play important role in some circumstances. For example, it has been known that, under the relaxation time approximation, for the previously studied rank-three spin conductivity, its extrinsic part can yield a nonzero spin Hall effect in systems that satisfies different symmetry requirement as the intrinsic part~\cite{Rafael2021,bose2021}.

Beyond the relaxation time approximation, the spin conductivity can be modified by the skew scattering and side jump contributions~\cite{Sinova2015}. Although the skew scattering contribution is typically of higher order in different scattering times, the side jump contribution is proven to be independent of the disorder density and determined by the form of the scattering potential alone, hence usually being mixed with the intrinsic contribution. For example, it has been shown that for some special scattering potential, the side jump contribution can  partially cancel the intrinsic part of the rank-three spin conductivity~\cite{xiao2021,Raimondi2012}, or fully cancel it to yield zero contribution at $\tau^0$~\cite{Raimondi2012}. In either cases, our work offers a fresh start from a rank-two intrinsic spin conductivity for future systematic study of the spin transport.

\begin{acknowledgments}
Y.G. acknowledges discussions with Qian Niu and Cong Xiao. D.H. and Y.G. acknowledges support from the Startup Foundation from USTC.
\end{acknowledgments}

%

\end{document}